\documentclass[two column, prA]{revtex4}%
\usepackage{amsmath}
\usepackage{graphicx}
\usepackage{amsfonts}
\usepackage{amssymb}%
\providecommand{\U}[1]{\protect\rule{.1in}{.1in}}

\begin{document}
\title{Lorentz invariant photon number density}
\author{Margaret Hawton}
\email{margaret.hawton@lakeheadu.ca}
\affiliation{Department of Physics, Lakehead University, Thunder Bay, ON, Canada, P7B 5E1}

\begin{abstract}
A Lorentz invariant positive definite expression for photon number density is
derived as the absolute square of the invariant scalar product of a
polarization sensitive position eigenvector and the photon wave function. It
is found that this scalar product is independent of the form chosen for the
wave function and that the normalized positive frequency vector
potential-electric field pair is a convenient choice of wave function in the
presence of matter. The number amplitude describing a localized state is a
$\delta$-function at the instant at which localization and detection are seen
as simultaneous.

\end{abstract}
\maketitle

\section{Introduction}

The concept of photon number density arises in the interpretation of
experiments such as photon counting and the creation of correlated photon
pairs in a nonlinear material.\ In spite of its relevance to experiment and
the foundations of quantum mechanics, no relativistically invariant positive
definite expression for photon number density exists at present. Mandel
\cite{Mandel66} defined a coarse-grained number operator to count photons in a
region large in comparison with the wavelength of any occupied mode of the
field. However photons can be counted, and hence localized, in a photodetector
much smaller than this wave length. Here I will show that the absolute square
of the Lorentz invariant scalar product of a localized state and the photon
wave function is a local positive definite number density.

In nonrelativistic quantum mechanics the $1$-particle number density is the
absolute square of the real space wave function. The wave function can be
obtained by projection of the state vector onto a basis of simultaneous
eigenvectors of the position and spin operators. Although this was long
thought to be impossible for the photon, we recently constructed a photon
position operator with commuting components
\cite{PositionOperator1,PositionOperator2} and derived a photon wave function
in this way \cite{WaveFunction1,WaveFunction2}. The Landau Peierls (LP) wave
function \cite{BB} and the positive frequency vector potential-electric field
(AE) wave function pair were considered and a scalar product was defined to
complete the Hilbert space. It was proved that the scalar product is invariant
under the similarity transformation relating the AE and LP forms of its
integrand, implying that they are equivalent when used for calculation of
transition amplitudes and expectation values.

Since position is an observable, the number amplitude to detect a particle at
position $\mathbf{r}$ with spin $\sigma$ equals the scalar product of the
corresponding position-spin eigenvector with the particle's wave function. In
real space this number amplitude, $\left\langle \mathbf{r},\sigma|\Psi\left(
t\right)  \right\rangle =\int d^{3}r^{\prime}\psi_{\mathbf{r},\sigma}^{\ast
}\left(  \mathbf{r}^{\prime}\right)  \Psi_{\sigma}\left(  \mathbf{r}^{\prime
},t\right)  $, is equal to the wave function, $\Psi_{\sigma}\left(
\mathbf{r},t\right)  $, only if the position eigenvectors are localized states
of the form $\psi_{\mathbf{r},\sigma}\left(  \mathbf{r}^{\prime}\right)
=\delta^{3}\left(  \mathbf{r}^{\prime}-\mathbf{r}\right)  .$ While this is
true for a nonrelativistic massive particle, it is not the case for the
photon. A transverse vector is not of the simple $\delta$-function form, and
thus it is not the amplitude to detect a photon at $\mathbf{r}$. In addition,
photons are often most conveniently described in terms of potentials or fields
because of their simple relationship to the matter current density. The
Fourier components of the vector potential and electric field describing a
localized state go as $\omega_{k}^{\mp1/2}$ where $\mathbf{k}$ is the wave
vector and $\omega_{k}$ is the angular frequency. They are not $\delta
$-functions even if their vector properties are ignored.

Since the boost operator just generates a change of point of view that does
not change the results of possible experiments \cite{Weinberg}, the scalar
product that predicts these results should be a relativistic invariant. Newton
and Wigner (NW) defined an invariant scalar product and a position eigenvector
at the origin that is invariant under rotation and inversion
\cite{NewtonWigner}. This and the displaced single particle states generated
in $\mathbf{k}$-space by the spatial translations $\exp\left(
-i\mathbf{k\cdot r}\right)  $ are orthonormal. Philips defined a
Lorentz-invariant localized wave function but lost the orthonormality
condition \cite{PhilipsGallardo}. The NW wave functions themselves are not
local in real space even in the simplest spin-zero case since they go as
$\omega_{k}^{1/2}$. However, as argued in the paragraph above, the number
amplitude is not in general equal to the real space wave function.

In this paper, the relationship between the photon wave function and number
density will be examined. In Section II, our previous work on the photon
position operator and wave function will be summarized. In Section III a
Lorentz invariant scalar product will be defined. The number amplitude for
arbitrary polarization, equal to the scalar product of a position eigenvector
and the wave function, will be found in Section IV. In Section V, interaction
with matter will be considered, and photon number density will be discussed in
relation to the recent and historical literature.

\section{Position operator and wave function}

The photon position operator with commuting components and transverse
eigenvectors can be written as \cite{PositionOperator2}
\begin{equation}
\widehat{\mathbf{r}}^{(\alpha,\chi)}=iD\left(  \omega_{k}\right)  ^{\alpha
}\nabla\left(  \omega_{k}\right)  ^{-\alpha}D^{-1} \label{PositionOperator}%
\end{equation}
where $\nabla$ is the $\mathbf{k}$-space gradient and $D=\exp\left(
-iS_{\mathbf{k}}\chi\right)  \exp\left(  -iS_{z}\phi\right)  \exp\left(
-iS_{y}\theta\right)  $ is the rotation matrix with Euler angles $\phi
,\theta,\chi.$ The operator $D$ rotates the unit vectors $\widehat{\mathbf{x}%
},$ $\widehat{\mathbf{y}}$ and $\widehat{\mathbf{z}}$ into the spherical polar
unit vectors $\widehat{\mathbf{\theta}},$ $\widehat{\mathbf{\phi}}$ and
$\widehat{\mathbf{k}}\mathbf{\ }$and then rotates $\widehat{\mathbf{\theta}}$
and $\widehat{\mathbf{\phi}}$ about $\mathbf{k}$ by $\chi(\mathbf{k})$. Thus
$\widehat{\mathbf{r}}^{(\alpha,\chi)}$ rotates the transverse and longitudinal
unit vectors to the fixed directions $\widehat{\mathbf{x}},$ $\widehat
{\mathbf{y}}$ and $\widehat{\mathbf{z}},$ eliminates the factor $\left(
\omega_{k}\right)  ^{\alpha},$ operates with $i\nabla$ to extract the position
information from the phase of the wave function, and then reinserts the factor
$\left(  \omega_{k}\right)  ^{\alpha}$\ and returns the unit vectors to their
original orientations. The eigenvectors of (\ref{PositionOperator}) form a
basis for the LP wave function if $\alpha=0,$ and for a wave function
proportional to the vector potential if $\alpha=-1/2$ and its conjugate
momentum if $\alpha=1/2.$

In the Schr\"{o}dinger picture (SP) the simultaneous eigenvectors of the
position and helicity operators,%
\begin{equation}
\mathbf{\psi}_{\mathbf{r},\sigma}^{(\alpha,\chi)}\left(  \mathbf{k}\right)
=\left(  \omega_{k}\right)  ^{\alpha}\frac{\exp\left(  -i\mathbf{k\cdot
r}\right)  }{\left(  2\pi\right)  ^{3/2}}\mathbf{e}_{\sigma}^{(\chi)}\text{,}~
\label{LocalizedStates}%
\end{equation}
form a basis for the Hilbert space, $\mathcal{H}$. The transverse unit vectors
are%
\begin{equation}
\mathbf{e}_{\mathbf{\sigma}}^{(\chi)}\left(  \mathbf{k}\right)  =\frac
{1}{\sqrt{2}}\left(  \widehat{\mathbf{\theta}}+i\sigma\widehat{\mathbf{\phi}%
}\right)  \exp\left(  -i\sigma\chi\right)  \label{e}%
\end{equation}
for circular polarizations $\sigma=\pm1$ and the longitudinal unit vector is
$\mathbf{e}_{0}=\widehat{\mathbf{k}}$. The real linear polarization unit
vectors%
\begin{align}
\mathbf{e}_{R1}^{(\chi)}  &  =\frac{1}{\sqrt{2}}\left(  \mathbf{e}_{1}%
^{(\chi)}+\mathbf{e}_{-1}^{(\chi)}\right)  ,\label{LinearPolarization}\\
\mathbf{e}_{R2}^{(\chi)}  &  =\frac{1}{i\sqrt{2}}\left(  \mathbf{e}_{1}%
^{(\chi)}-\mathbf{e}_{-1}^{(\chi)}\right)  ,\nonumber
\end{align}
give eigenvectors of $\widehat{\mathbf{r}}^{(\alpha,\chi)}$ but not the
helicity operator. While all $\chi$ are needed to interpret an experiment that
measures polarization, only one basis is needed to describe the photon state,
and the $\chi=0$ definite helicity basis will be used here for simplicity. In
this basis, the probability amplitude for the state with polarization
$\mathbf{e}_{\mathbf{\sigma}}^{(\chi)}$ incorporates the phase factor
$\exp\left(  -i\sigma\chi\right)  $.

The wave function was obtained in \cite{WaveFunction1} as the projection of
the quantum electrodynamic (QED) state vector onto a basis of
position-helicity eigenvectors. If the operator $a_{\sigma}^{\dagger}\left(
\mathbf{k}\right)  $ creates a photon with wave vector $\mathbf{k}$ and
circular polarization $\mathbf{e}_{\sigma}^{(0)}$, the $1$-photon $\mathbf{k}%
$-space basis states are $\left\vert \mathbf{k},\sigma\right\rangle
=a_{\sigma}^{\dagger}\left(  \mathbf{k}\right)  \left\vert 0\right\rangle $
where $\left\vert 0\right\rangle $ is the vacuum state. Creation operators for
a photon at position $\mathbf{r}$ can be defined as%
\begin{equation}
\widehat{\mathbf{\psi}}_{\mathbf{r},\sigma}^{(\alpha)\dagger}=\left(
\omega_{k}\right)  ^{\alpha}\frac{\exp\left(  -i\mathbf{k\cdot r}\right)
}{\left(  2\pi\right)  ^{3/2}}\mathbf{e}_{\sigma}^{(0)\ast}\left(
\mathbf{k}\right)  a_{\sigma}^{\dagger}\left(  \mathbf{k}\right)  .
\label{CreationOperator}%
\end{equation}
Any state vector can be expanded in Fock space as $\left\vert \Psi\left(
t\right)  \right\rangle =\sum_{n=0}^{\infty}c_{n}\left\vert \Psi_{n}\left(
t\right)  \right\rangle $ where the $1$-photon term $\left\vert \Psi
_{1}\left(  t\right)  \right\rangle =\sum_{\sigma}\int d^{3}k\left\langle
\mathbf{k},\sigma|\Psi\left(  t\right)  \right\rangle \left\vert
\mathbf{k},\sigma\right\rangle $ is completely described by the probability
amplitude
\begin{equation}
\left\langle \mathbf{k},\sigma|\Psi\left(  t\right)  \right\rangle \equiv
c_{\sigma}\left(  \mathbf{k}\right)  \exp\left(  -i\omega_{k}t\right)  .
\label{QEDProbabilityAmplitude}%
\end{equation}
The projection onto the definite helicity position eigenvectors,
$\mathbf{\Psi}_{\sigma}^{(\alpha)}\left(  \mathbf{r},t\right)  =\left\langle
0\left\vert \widehat{\mathbf{\psi}}_{\mathbf{r},\sigma}^{(\alpha)}\right\vert
\Psi\left(  t\right)  \right\rangle $, is a six component wave function whose
dynamics is described by a diagonal Hamiltonian \cite{BB,WaveFunction1}. The
expectation value can be evaluated using $\left[  a_{\sigma}\left(
\mathbf{k}\right)  ,a_{\sigma^{\prime}}\left(  \mathbf{k}^{\prime}\right)
\right]  =\delta_{\sigma,\sigma^{\prime}}\delta^{3}\left(  \mathbf{k}%
-\mathbf{k}^{\prime}\right)  $. Here the focus is on the scalar product and
polarization will be summed over to give the $3$-vector wave function
$\mathbf{\Psi}^{(\alpha)}\left(  \mathbf{r},t\right)  $ where
\begin{align}
\mathbf{\Psi}_{\sigma}^{(\alpha)}\left(  \mathbf{r},t\right)   &  =\int
d^{3}k\ c_{\sigma}\left(  \mathbf{k}\right) \label{WaveFunction}\\
&  \times\mathbf{e}_{\sigma}^{(0)}\left(  \mathbf{k}\right)  \left(
\omega_{k}\right)  ^{\alpha}\frac{\exp\left(  i\mathbf{k\cdot r}-i\omega
_{k}t\right)  }{\left(  2\pi\right)  ^{3/2}},\nonumber\\
\mathbf{\Psi}^{(\alpha)}\left(  \mathbf{r},t\right)   &  =\sum_{\sigma=\pm
1}\mathbf{\Psi}_{\sigma}^{(\alpha)}\left(  \mathbf{r},t\right)  .\nonumber
\end{align}
The $\mathbf{k}$-space wave function will be defined as%
\begin{align}
\mathbf{\Psi}_{\sigma}^{(\alpha)}\left(  \mathbf{k}\right)   &  =c_{\sigma
}\left(  \mathbf{k}\right)  \mathbf{e}_{\sigma}^{(0)}\left(  \mathbf{k}%
\right)  \left(  \omega_{k}\right)  ^{\alpha},\label{kWaveFunction}\\
\mathbf{\Psi}^{(\alpha)}\left(  \mathbf{k}\right)   &  =\sum_{\sigma=\pm
1}\mathbf{\Psi}_{\sigma}^{(\alpha)}\left(  \mathbf{k}\right)  .\nonumber
\end{align}

\section{Invariant scalar product}

To prove the invariance of the scalar product that completes $\mathcal{H}$,
$4$-vector notation and the Lorentz gauge will be used. The contravariant
energy-momentum $4$-vector is $\hbar k$ where $k=k^{\nu}=(\omega
_{k}/c,\mathbf{k})$ and the covariant $4$-vector is $k_{\mu}=g_{\mu\nu}k^{\nu
}$ where $g$ is a diagonal tensor with $g_{00}=-1$ and $g_{ii}=1$ for $i=1$ to
$3.$ In addition to the transverse and longitudinal eigenvectors, scalar
position eigenvectors can be defined \cite{WaveFunction99} and the
$4$-potential can be used as wave function \cite{Gross}. In the Lorentz gauge
the momentum conjugate to $A^{(+)\mu}$ is $\Pi^{(+)\mu}=\epsilon_{0}\partial
A^{(+)\mu}/\partial t$ \cite{CohenTannoudji} where the superscript $(+)$
denotes the positive frequency part. This is equivalent to $\mathbf{\psi
}^{(1/2)}=i\partial\mathbf{\psi}^{(-1/2)}/\partial t$ that follows from
(\ref{WaveFunction}). Since only positive frequencies arise in the wave
function, the invariant volume integral can be written as%
\begin{equation}
\int d^{4}k\delta\left(  k^{\mu}k_{\mu}+\kappa^{2}\right)  \Theta\left(
\omega_{k}\right)  =c\int\frac{d^{3}k}{2\omega_{k}} \label{InvariantIntegral}%
\end{equation}
where the Heaviside step function $\Theta\left(  \omega_{k}\right)  $ is
invariant under proper Lorentz transformations and $\kappa=mc/\hbar=0$ for a
free photon.

The real space wave function satisfying $\Psi^{(-1/2)}\left(  x\right)
=\sqrt{2\epsilon_{0}/\hbar}A^{(+)}\left(  x\right)  $ \cite{WaveFunction1} is
a $4$-vector. Use of (\ref{kWaveFunction}) in (\ref{WaveFunction}) for
$\alpha=-1/2$ gives%
\begin{equation}
\Psi^{(-1/2)}\left(  x\right)  =\int\frac{d^{3}k}{\omega_{k}}\frac{\exp\left(
ikx\right)  }{\left(  2\pi\right)  ^{3/2}}\Psi^{(1/2)}\left(  k\right)
\label{FT}%
\end{equation}
for physical states with $\exp\left(  -i\omega_{k}t\right)  $ time dependence.
This implies that $\Psi_{\sigma}^{(1/2)}\left(  k\right)  $ must be a
$4$-vector. Thus the scalar product of $\left\vert \Phi_{1}\right\rangle $ and
$\left\vert \Psi_{1}\right\rangle ,$
\begin{equation}
\left\langle \Phi_{1}|\Psi_{1}\right\rangle =\int\frac{d^{3}k}{\omega_{k}}%
\Phi^{(1/2)\mu\ast}\left(  k\right)  \Psi_{\mu}^{(1/2)}\left(  k\right)  ,
\label{InvariantScalarProduct}%
\end{equation}
is an invariant. This is the form normally used in field theory, incorporating
the metric $\omega_{k}^{-1}$. Alternatively, the integrand $\Phi^{(1/2)\mu
\ast}\left(  k\right)  \Psi_{\mu}^{(1/2)}\left(  k\right)  /\omega_{k}$ can be
written as the product of LP wave functions, $\Phi^{(0)\mu\ast}\left(
k\right)  \Psi_{\mu}^{(0)}\left(  k\right)  $, or as the potential-conjugate
momentum product, $\Phi^{(-1/2)\mu\ast}\left(  k\right)  \Psi_{\mu}%
^{(1/2)}\left(  k\right)  $, to give the invariant scalar product
(\ref{InvariantScalarProduct}) as%
\begin{equation}
\left\langle \Phi_{1}|\Psi_{1}\right\rangle =\int d^{3}k\Phi^{(-\alpha)\mu
\ast}\left(  k\right)  \Psi_{\mu}^{(\alpha)}\left(  k\right)  .
\label{PotentialScalarProduct}%
\end{equation}
In the Lorentz gauge there are longitudinal and scalar photons in addition to
the observable transverse photons. Scalar and longitudinal photons can be
dealt with by introducing an indefinite metric \cite{CohenTannoudji}. However
the Lorentz gauge was only needed to prove invariance and its further use here
is not required.

In any specific reference frame it is possible to make a gauge transformation
to the transverse (Coulomb) gauge. Eq. (\ref{PotentialScalarProduct}) then
reduces to%
\begin{equation}
\left\langle \Phi_{1}|\Psi_{1}\right\rangle =\int d^{3}k\mathbf{\Phi
}^{(-\alpha)\ast}\left(  \mathbf{k}\right)  \cdot\mathbf{\Psi}^{(\alpha
)}\left(  \mathbf{k}\right)  . \label{TransverseScalarProduct}%
\end{equation}
The conjugate momentum is $\mathbf{\Pi}^{(+)}\mathbf{=}-\epsilon_{0}%
\mathbf{E}_{\perp}^{(+)}$ for the minimal coupling Hamiltonian ($\perp$
denotes the transverse part), so the $\alpha=\mp1/2$ conjugate pair are the
vector potential and the electric field (AE). Since the $\left(  \omega
_{k}\right)  ^{\pm\alpha}$ factors cancel and the dot product of like unit
vectors is unity, Eq. (\ref{TransverseScalarProduct}) can also be written as%
\begin{equation}
\left\langle \Phi_{1}|\Psi_{1}\right\rangle =\sum_{\sigma=\pm1}\int
d^{3}kd_{\sigma}^{\ast}\left(  \mathbf{k}\right)  c_{\sigma}\left(
\mathbf{k}\right)  \label{ScalarProduct}%
\end{equation}
where the $d^{\prime}s$ are the coefficients in the expansion of $\left\vert
\Phi_{1}\right\rangle .$ This is just the QED scalar product.

For any operator, $\widehat{O}$, describing an observable, $\widehat{O}%
\Psi_{\sigma}^{(\alpha)}\left(  \mathbf{k}\right)  $ must be in $\mathcal{H}$
and the eigenvalues of $\widehat{O}$ must be real. This is obviously the case
for the momentum and energy operators, $\hbar\mathbf{k}$ and $\hbar\omega
_{k}.$ For the position operator, using (\ref{kWaveFunction}),%
\begin{equation}
\widehat{\mathbf{r}}^{(\alpha,\chi)}\Psi_{\sigma}^{(\alpha)}\left(
\mathbf{k}\right)  =\left(  \omega_{k}\right)  ^{\alpha}\mathbf{e}_{\sigma
}^{(\chi)}\left(  \mathbf{k}\right)  i\nabla\left[  c_{\sigma}\left(
\mathbf{k},t\right)  \exp\left(  i\sigma\chi\right)  \right]  \label{ropPsi}%
\end{equation}
since $\left(  \omega_{k}\right)  ^{\alpha}\mathbf{e}_{\sigma}^{(\chi)}\left(
\mathbf{k}\right)  $ is an eigenvector of $\widehat{\mathbf{r}}^{(\alpha
,\chi)}$ with eigenvalue zero. Thus (\ref{ropPsi}) is in $\mathcal{H}$. Using
the scalar product (\ref{InvariantScalarProduct}) the wave function is just
$\Psi^{(1/2)\mu}\left(  k\right)  $ and it can be proved as in
\cite{PikeSarkar} that $\widehat{\mathbf{r}}^{(1/2,\chi)}$ is Hermitian. If
(\ref{TransverseScalarProduct}) is used, it can be proved using (\ref{ropPsi})
that $\left\langle \widehat{\mathbf{r}}^{(-\alpha,\chi)}\Phi_{1}|\Psi
_{1}\right\rangle =\left\langle \Phi_{1}|\widehat{\mathbf{r}}^{(\alpha,\chi
)}\Psi_{1}\right\rangle ^{\ast}$ which implies that $\widehat{\mathbf{r}%
}^{(0,\chi)}$ is Hermitian and must have real eigenvalues and orthonormal
eigenvectors. This follows the usual rules of operator algebra since $D$ is
unitary and $i\nabla$ is Hermitian. The $\alpha=\pm1/2$ basis is related to
the $\alpha=0$ basis by $\widehat{\mathbf{r}}^{(\alpha,\chi)}=\left(
\omega_{k}\right)  ^{\alpha}\widehat{\mathbf{r}}^{(0,\chi)}\left(  \omega
_{k}\right)  ^{-\alpha}$. This similarity transformation preserves the scalar
product and hence expectation values and the reality of the position
eigenvectors \cite{WaveFunction1,Mostaf02}. The wave functions $\mathbf{\Psi
}_{\sigma}^{(-1/2)}$ and $\mathbf{\Psi}_{\sigma}^{(1/2)}$ form a biorthonormal
pair and the position operator satisfies $\widehat{\mathbf{r}}^{(-1/2,\chi
)\dagger}=\widehat{\mathbf{r}}^{(1/2,\chi)}$, so it can be called
pseudo-Hermitian \cite{Mostaf02}.

In real space, the most convenient form for the scalar product is not obvious.
If (\ref{InvariantScalarProduct}) is transformed directly, the metric factor
$\omega_{k}^{-1}$ must be replaced by the inverse of the Hamiltonian operator,
implying a nonlocal integrand \cite{BB}. Eq. (\ref{TransverseScalarProduct})
transformed to real space is%
\begin{equation}
\left\langle \Phi_{1}|\Psi_{1}\right\rangle \ =\int d^{3}r\mathbf{\Phi
}^{(-\alpha)\ast}\left(  \mathbf{r},t\right)  \cdot\mathbf{\Psi}^{(\alpha
)}\left(  \mathbf{r},t\right)  , \label{rSpaceScalarProduct}%
\end{equation}
as can be verified by substitution of (\ref{WaveFunction}) and integration
over $d^{3}r$ to give $\delta^{3}\left(  \mathbf{k}-\mathbf{k}^{\prime
}\right)  $ and then (\ref{TransverseScalarProduct}). The integrand of
(\ref{rSpaceScalarProduct}) is local, making it a useful form of the scalar product.

It can be seen by inspection of Eqs. (\ref{InvariantScalarProduct}) to
(\ref{rSpaceScalarProduct}) that the scalar product is unaffected by the
change of metric from (\ref{InvariantScalarProduct}) to
(\ref{PotentialScalarProduct}), and is invariant under the similarity
transformations between the AE\ and LP forms of the wave function and under
the unitary transformation between $\mathbf{r}$-space and $\mathbf{k}$-space.
It is also invariant under the unitary transformations $\exp\left(
-iS_{\mathbf{k}}\chi\right)  $\ to $\chi\neq0$ bases. The AE form of the wave
function is preferable in most applications, since the relationship of the LP
wave function to matter source terms is nonlocal \cite{Cook}, but the choice
is a matter of convenience.

\section{Photon number amplitude}

The probability amplitude for a measured result is the amplitude for a
transition from the photon state described by $c_{\sigma}\left(
\mathbf{k}\right)  $ to a final state that is an eigenvector of the operators
representing the experiment. For a measurement of momentum $\hbar\mathbf{k}$
and polarization $\mathbf{e}_{\sigma}^{(\chi)}$\ this final state is described
by $d_{\sigma^{\prime}}\left(  \mathbf{k}^{\prime}\right)  $ $=\delta
_{\sigma^{\prime},\sigma}\exp\left(  -i\sigma\chi\right)  \delta^{3}\left(
\mathbf{k}^{\prime}-\mathbf{k}\right)  $. Substitution in (\ref{ScalarProduct}%
) gives the probability amplitude $c_{\sigma}\left(  \mathbf{k}\right)
\exp\left(  i\sigma\chi\right)  $ where the factor $\exp\left(  i\sigma
\chi\right)  $ rotates the polarization about $\mathbf{k}$ by $-\chi$. The
helicity is an invariant and a Lorentz transformation just changes $\chi$
\cite{Weinberg}, so the probability to detect a photon with definite helicity
and wave vector is invariant. For a measurement of the linear polarization
$\mathbf{e}_{Rj}^{(\chi)}$, given by (\ref{LinearPolarization}), the momentum
eigenvectors are $d_{\sigma}\left(  \mathbf{k}^{\prime}\right)  $
$=\exp\left(  -i\sigma\chi\right)  \delta^{3}\left(  \mathbf{k}^{\prime
}-\mathbf{k}\right)  /\sqrt{2}$ and $d_{\sigma}\left(  \mathbf{k}^{\prime
}\right)  $ $=i\left(  -1\right)  ^{\sigma}\exp\left(  -i\sigma\chi\right)
\delta^{3}\left(  \mathbf{k}^{\prime}-\mathbf{k}\right)  /\sqrt{2}$ for $j=1$
and $2$ respectively,\ so (\ref{ScalarProduct}) gives the probability
amplitude to detect a photon with wave vector $\mathbf{k}$ and this
polarization direction as
\begin{align}
c_{R1}\left(  \mathbf{k},\chi\right)   &  =\frac{1}{\sqrt{2}}\left[
c_{1}\left(  \mathbf{k}\right)  \exp\left(  i\chi\right)  +c_{-1}\left(
\mathbf{k}\right)  \exp\left(  -i\chi\right)  \right]  ,\label{cL}\\
c_{R2}\left(  \mathbf{k},\chi\right)   &  =\frac{i}{\sqrt{2}}\left[
c_{1}\left(  \mathbf{k}\right)  \exp\left(  i\chi\right)  -c_{-1}\left(
\mathbf{k}\right)  \exp\left(  -i\chi\right)  \right]  .\nonumber
\end{align}
If $\chi=0$ the measured polarization directions are $\widehat{\mathbf{\theta
}}$ and $\widehat{\mathbf{\phi}}$, and $c_{R1}$ and $c_{R2}$ are the
amplitudes for transverse magnetic and transverse electric modes respectively
\cite{Hacyan}. Probability density is the absolute valued squared, so the
probability to detect a photon with definite helicity is $\chi$-independent,
but a linear polarization measurement is sensitive to phase. For $c_{\sigma
}\propto\exp\left(  -i\sigma\chi^{\prime}\right)  $, $c_{R1}\propto\cos
\Delta\chi$ and $c_{L2}\propto\sin\Delta\chi$ where $\Delta\chi=\chi^{\prime
}-\chi$ is the polarization angle of the photon relative to the polarization
measured by the apparatus.

The physical states with circular polarization $\mathbf{e}_{\sigma}^{(\chi)}$
localized at position $\mathbf{r}$ at fixed time $t$ have the $\mathbf{k}%
$-space amplitudes
\begin{equation}
d_{\sigma^{\prime}}\left(  \mathbf{k,r},t,\chi\right)  =\delta_{\sigma
^{\prime},\sigma}\frac{\exp\left(  -i\sigma\chi-i\mathbf{k}\cdot
\mathbf{r}+i\omega_{k}t\right)  }{\left(  2\pi\right)  ^{3/2}}%
.\label{PhysicalLocalized}%
\end{equation}
This is the probability amplitude for a position eigenvector if the phase
$\chi\rightarrow\chi-\sigma\omega_{k}t$ in (\ref{PositionOperator}). The
amplitude for a photon in state $c_{\sigma}\left(  \mathbf{k}\right)  $ to
make a transition to this state is given by substitution of
(\ref{PhysicalLocalized} ) into (\ref{ScalarProduct}) as%
\begin{equation}
c_{\sigma}\left(  \mathbf{r},t\right)  =\int d^{3}k\frac{\exp\left(
i\sigma\chi+i\mathbf{k}\cdot\mathbf{r}-i\omega_{k}t\right)  }{\left(
2\pi\right)  ^{3/2}}c_{\sigma}\left(  \mathbf{k}\right)
.\label{PhotonNumberAmplitude}%
\end{equation}
It equals the inverse Fourier transform of $c_{\sigma}\left(  \mathbf{k}%
\right)  \exp\left(  i\sigma\chi-i\omega_{k}t\right)  $.

As a consistency check, consider a photon with polarization $\mathbf{e}%
_{\sigma}^{(\chi)}$ localized at $\mathbf{r}^{\prime}$ at time $t^{\prime}$.
The probability amplitude for this state is $c_{\sigma}\left(  \mathbf{k}%
\right)  =d_{\sigma}\left(  \mathbf{k,r}^{\prime},t^{\prime},\chi\right)  $
and (\ref{PhotonNumberAmplitude}) can be integrated to give the number
amplitude $c_{\sigma}\left(  \mathbf{r},t\right)  =\delta_{\sigma
,\sigma^{\prime}}\delta^{3}\left(  \mathbf{r}-\mathbf{r}^{\prime}\right)  $ if
$t=t^{\prime}$, that is if localization and measurement are seen as
simultaneous. All $\mathbf{k}^{\prime}s$ are included with equal weight in
(\ref{PhysicalLocalized}) and exact localization, which is limited by the
Hegerfeldt theorem \cite{Hegerfeldt}, is only possible because of interference
effects between the converging and the diverging wave. Eq.
(\ref{PhysicalLocalized}) describes a situation where there is no physical
detector at $\mathbf{r}$,\ so absorption is followed immediately by re-emission.

\section{Conclusion}

The most widely accepted photon wave function in the current literature is the
Glauber photodetection amplitude \cite{Scully}, $\mathbf{E}_{\perp}%
^{(+)}\left(  \mathbf{r},t\right)  =\left\langle 0\left\vert \widehat
{\mathbf{E}}\left(  \mathbf{r},t\right)  \right\vert \Psi\right\rangle $,
where $\widehat{\mathbf{E}}\left(  \mathbf{r},t\right)  $ is the Heisenberg
picture (HP) electric field operator and $\mathbf{E}_{\perp}^{(+)}$ is
proportional to $\mathbf{\Psi}^{(1/2)}$ given by (\ref{WaveFunction}). This
can be combined with $\mathbf{\Psi}^{(-1/2)}$ to give a scalar product with a
local integrand and a complete description of photon quantum mechanics. Since
the scalar product is invariant under unitary and similarity transformations
and the prediction of the experimental results requires only the scalar
product, the choice of wave function can be based on convenience. If desired,
the remaining fields $\mathbf{D}^{(+)}=\epsilon_{0}\mathbf{E}^{(+)}%
+\mathbf{P}^{(+)}$, $\mathbf{H}^{(+)}$ and $\mathbf{B}^{(+)}=\mu_{0}%
\mathbf{H}^{(+)}+\mathbf{M}^{(+)}$ can defined similarly
\cite{Sipe,WaveFunction1} where $\left\vert 0\right\rangle $ is the
matter-field ground state and $\widehat{\mathbf{P}}$ and $\widehat{\mathbf{M}%
}$ create and destroy matter excitations. Since the HP field operators satisfy
classical dynamical equations, the equations satisfied by these positive
frequency fields are identical in form to Maxwell's equations and describe the
dynamics of the $1$-polariton state.

Evaluation of the photon number amplitude (\ref{PhotonNumberAmplitude}) is
straightforward since it just requires integration of the \emph{scalar} QED
probability amplitude. Polarization unit vectors as a function of $\mathbf{k}$
can be selected for convenience, for example, $\chi=-\phi$ gives
$\mathbf{e}_{\mathbf{\sigma}}^{(-\phi)}\left(  k\widehat{\mathbf{z}}\right)
=\left(  \widehat{\mathbf{x}}+i\sigma\widehat{\mathbf{y}}\right)  /\sqrt{2}$
for a paraxial beam propagating parallel to $\widehat{\mathbf{z}}$. Number
amplitude itself has intrinsic physical significance. It is number amplitude
that, after Schmidt diagonalization, was used as a photon wave function by
Chan, Law and Eberly \cite{Eberly}. Also, number amplitude is proportional to
$\mathbf{E}_{\perp}^{(+)}$ to a good approximation. The positive frequency
electric field of a localized photon state drops off as $r^{-7/2}$ where $r$
is the distance from the point of photon localization \cite{Amrein}. Thus
photon number density, equal to the absolute square of
(\ref{PhotonNumberAmplitude}), is an excellent approximation to the Glauber
photodetection probability, proportional to $\left\vert \mathbf{E}_{\perp
}^{(+)}\right\vert ^{2}$.

Number density may prove to be of fundamental importance. Cook pointed out
that photon current density cannot be made precise within the Glauber theory,
so that it fails to provide a complete description of photon transport
\cite{Cook}. Calculation of the probability density to detect a photon as the
absolute square of the scalar product of the $1$-photon wave function and a
position eigenvector using (\ref{PhotonNumberAmplitude}) follows the usual
rules of quantum mechanics. There is no such connection between Glauber
photodetection theory and the basic rules of quantum mechanics. It is an
ongoing goal of this author to thoroughly understand the relationship of
photon number density to photon fields, energy density, the interpretation of
photon counting experiments, and fundamental issues relating to photon localizability.

In summary, photon number amplitude in real space was calculated as the
invariant scalar product of a localized state with definite polarization and
the $1$-photon wave function. It gives an excellent approximation to the
Glauber photodetection probability, with the advantage that the orthonormal
position eigenvectors lead to mutually exclusive probability densities and an
integrated probability of unity. When described in terms of this number
amplitude, the localized basis states combine the orthonormality of the NW
states with the Lorentz invariance of Philips' localized states
\cite{NewtonWigner,PhilipsGallardo}. The positive frequency vector potential
and electric field provide a complete description of photon quantum mechanics,
including interaction with matter. This should make it possible to test the
relevance of the photon number concept to experiment.

\ \textit{Acknowledgements: }The author thanks the Natural Sciences and
Engineering Research Council for financial support.

\end{document}